\documentclass[12pt,twoside,a4paper]{article} 
\usepackage{amsmath,amssymb,latexsym,theorem,bbm,shapepar,natbib,epsfig}
\newfont{\msa}{msam10 scaled\magstep1}
\newfont{\ssmsa}{msam9}

\def\crps{\mathop{\hbox{\rm crps}}}

\numberwithin{equation}{section}

\title{Probabilistic wind speed forecasting using Bayesian model
  averaging with truncated normal components}

\author{{\sc S\'andor Baran}\\ 
         Faculty of Informatics, University of Debrecen\\
         Kassai \'ut 26, H-4028 Debrecen, Hungary \\ 
        and \\
        Institute of Applied Mathematics, University of Heidelberg\\
        Im Neuenheimer Feld 294, D-69120 Heidelberg, Germany
     }

\date{}

\begin{document}
\pagestyle{myheadings}

\maketitle

\begin{abstract}
Bayesian model averaging (BMA) is a statistical method for post-processing 
forecast ensembles of atmospheric variables,
obtained from  multiple runs of numerical weather prediction  models,
in order to create calibrated predictive probability
density functions (PDFs). The BMA predictive PDF of the future weather
quantity is the mixture of the individual PDFs corresponding to the
ensemble members and the weights and model parameters are
estimated using ensemble members and validating observation from
a given training period.

In the present paper we introduce a BMA model for calibrating wind
speed forecasts, where the components PDFs follow truncated normal distribution
with cut-off at zero, and apply it to the ALADIN-HUNEPS
ensemble of the Hungarian Meteorological Service. Three
parameter estimation methods are proposed and each of the
corresponding models outperforms the 
traditional gamma BMA model both in calibration and in accuracy of
predictions. 
Moreover, since here the  maximum likelihood 
estimation of the parameters does not require numerical optimization,
modelling can be performed much faster than in case of gamma mixtures.

\end{abstract}

\section{Introduction}
   \label{sec:sec1} 

The most important aim of weather forecasting is to give a robust and
reliable prediction of the future state of the atmosphere based on
observational data, prior forecasts and mathematical models describing
the dynamical and physical behaviour of the atmosphere. 
These models consists of sets of  hydro-thermodynamic
non-linear partial differential equations of the atmosphere and its
coupled systems (like surface or oceans for instance) and have only
numerical solutions. The difficulty
with these numerical weather prediction models is that since the
atmosphere has a chaotic character the solutions  
strongly depend on the initial conditions and also on other uncertainties 
related to the numerical weather prediction process. Therefore, 
the results of such models are never fully accurate.        
A possible solution is to run the model with different
initial conditions (since the uncertainties in the initial conditions
are one of the most important sources of uncertainty) and produce an
ensemble of forecasts. Using a forecast ensemble one can estimate the
probability distribution of future weather variables which allows
probabilistic weather forecasting \citep{gr}, where not only the
future atmospheric states are 
predicted, but also the related uncertainty information. 
The ensemble
prediction method was proposed by \citet{leith} and since its first
operational implementation \citep{btmp,tk} it has become a widely used
technique all over the world and the users understand more and more
its merits and economic value as well. However, although e.g. the ensemble mean 
on average yields better forecasts of a meteorological quantity than
any of the individual ensemble members, it is often the case that the
ensemble is under-dispersive 
and in this way, uncalibrated \citep{bhtp}, so that calibration is
needed to account for this deficiency.  

The Bayesian model averaging (BMA) method for
post-processing ensembles in order to calibrate them was introduced by
\citet{rgbp}. The basic idea of BMA is that for each member of the ensemble
forecast there is a corresponding  conditional probability
density function (PDF) that can be interpreted as the
conditional PDF of the future weather quantity provided the considered
forecast is the best one. Then the BMA predictive PDF of the future
weather quantity is the weighted sum of the individual PDFs
corresponding to the ensemble members and the weights are based on the
relative performance of the ensemble members during a given training
period. In this way BMA is a special, fixed parameter version of 
dynamic model averaging method developed by \citet{rke}. The weight
parameters and the parameters of the individual 
PDFs are estimated using linear regression and maximum likelihood (ML)
method, where the maximum of the likelihood function is found by EM
algorithm. In practice, the performance of the individual
ensemble members should have a clear characteristic (and not a random
one) or if it is not the case this fact should be taken into account
at the calibration process \citep[see e.g.][]{frg}.  In \citet{rgbp} the BMA
method was successfully applied to obtain 48 hour forecasts of surface
temperature and sea level pressure in the North American Pacific
Northwest based on the 5 members of the University of Washington
Mesoscale Ensemble \citep{gm}. These weather quantities can be
modeled by normal distributions, so the predictive PDF is a Gaussian mixture.
Later, \citet{srgf} developed a discrete-continuous BMA model for
precipitation 
forecasting, where the discrete part corresponds to the event of no
precipitation, while the cubic root of the precipitation amount
(if it is positive) is modeled by a gamma distribution. In
\citet{sgr10} the BMA method was used for wind speed forecasting and 
the component PDFs follow gamma distributions, while using von Mises
distribution to model angular data, 
\citet{bgrgg} introduced a BMA scheme to predict surface wind
direction. Finally, \citet{sgr12} described a BMA model for
wind vector forecasting, where wind vectors are modeled using
bivariate normal distribution. 

The bivariate normal model for wind vectors is also used in the
ensemble model output statistics (EMOS) method for post-processing
ensemble forecasts \citep{stg}. The EMOS, introduced by \citet{grwg}
for calibrating ensemble forecasts following normal distribution (sea
level pressure, temperature), produces a single normal PDF, where the
mean and the variance depends on the ensemble members. However, the
method can be extended to truncated normal distribution \citep{tg}, too, and
in this way it can be used for calibrating wind speed data.

In the present paper we develop a BMA model for wind speed forecasting where
the component PDFs, similarly to the EMOS PDF of \citet{tg}, follow
truncated normal distribution. The advantage of this model to the
gamma model of \citet{sgr10} is that for  parameter estimation
the truncated data EM algorithm for Gaussian mixture models 
\citep{ls12} can be used which works with closed formulae both in
expectation (E) and in maximization (M) steps. In this way the
estimation of parameters is much faster, which is a key issue in operational
applications. 

We test our model on the ensemble forecasts of wind speed produced by
the operational Limited Area Model Ensemble  
Prediction System (LAMEPS) of the Hungarian Meteorological Service
(HMS) called ALADIN-HUNEPS \citep{hagel, horanyi} and compare the
results obtained with the corresponding results of \citet{bhn1} where
for calibration the BMA gamma model of \citet{sgr10} was used.

\section{Data}
  \label{sec:sec2}

The ALADIN-HUNEPS system of the HMS covers a large part of Continental
Europe with a 
horizontal resolution 
of 12 km and it is obtained by dynamical downscaling (by the ALADIN
limited area model) of the global
ARPEGE based PEARP system of M\'et\'eo France \citep{hkkr,dljn}. The
ensemble consists of 11 members, 10 initialized from perturbed initial
conditions and one control member from the unperturbed analysis,
implying that the ensemble contains groups of exchangeable
forecasts. The data base contains 11 member ensembles of 42 hour 
forecasts for 10 meter wind speed (given in m/s)
for 10 major cities in 
Hungary (Miskolc, Szombathely, Gy\H or, Budapest, Debrecen, Ny\'\i regyh\'aza,
Nagykanizsa, P\'ecs, Kecskem\'et, Szeged) produced by the
ALADIN-HUNEPS system of the HMS, together with the corresponding
validating observations for the period between October 1, 2010 and March 25,
2011 (176 days, or 1760 data points). The forecasts are
initialized at 18 UTC. The data set is fairly complete since there are
only two days 
(18.10.2010 and 15.02.2011) where three ensemble members are missing
for all sites and one day (20.11.2010) when no forecasts are available. 

\begin{figure}[t]
\begin{center}
\leavevmode
\epsfig{file=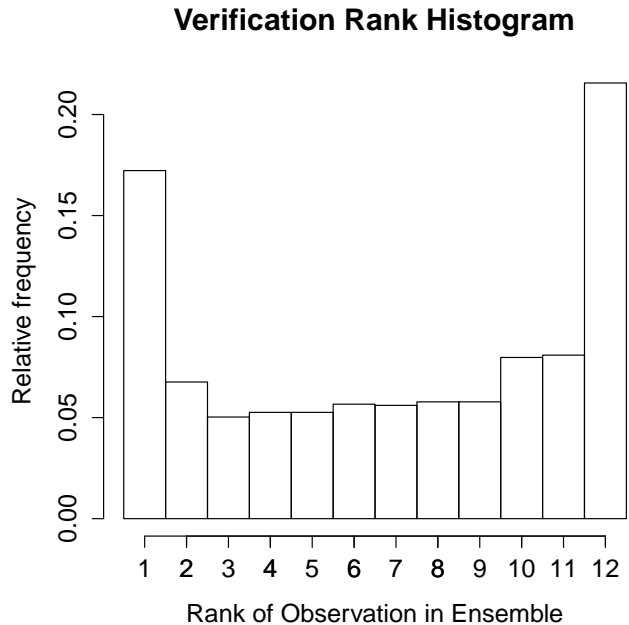,height=9cm, angle=-90}
\caption{Verification rank histogram of the 11-member ALADIN-HUNEPS
  ensemble. Period: October 1, 2010 -- March 25, 2011.} 
\label{fig:fig1}
\end{center}
\end{figure}
Figure \ref{fig:fig1} shows the verification rank histogram of the raw
ensemble. This is the histogram of ranks of validating
observations with respect to the corresponding ensemble
forecasts computed from the ranks at all stations and over the whole
  verification period \citep[see e.g.][Section 7.7.2]{wilks2}. 
This histogram is 
far from the desired uniform distribution as in many cases the
ensemble members either underestimate or overestimate the validating
observations (the ensemble range contains the observed wind speed 
in $61.21\%$ of the cases, while its nominal value equals $10/12$, i.e
$83.33 \%$). Hence, the ensemble is under-dispersive
and in this way it is uncalibrated. Therefore, statistical
post-processing is required to improve the forecasted probability
density function.

\section{Methods}
  \label{sec:sec3}
\subsection{Bayesian model averaging}
  \label{subs:subs3.1}

Let \ $f_1,f_2,\ldots ,f_M$ \ denote the ensemble forecast of a
certain weather quantity \ $X$ \ for a given location and time.
In BMA for ensemble forecasting \citep{rgbp} to each ensemble member \
$f_k$ \ corresponds a component PDF \ $g_k(x \vert f_k, \theta_k)$, \
where \ $\theta_k$ \ is a parameter to be estimated. The BMA
predictive PDF of \ $X$ \ is
\begin{equation}
  \label{eq:eq3.1}
p(x\vert\, f_1, \ldots ,f_M;\theta_1, \ldots
,\theta_M):=\sum_{k=1}^M\omega _k g_k(x \vert\, f_k, \theta_k),
\end{equation}
where the weight \ $\omega_k$ \ is connected to the relative
performance of the ensemble member \ $f_k$ \ during the training
period. Obviously, the weights form a probability distribution and in
this way they are nonnegative and \ $\sum_{k=1}^M \omega_k=1$. \ 

BMA model \eqref{eq:eq3.1} is valid only in the cases when the sources
of the ensemble members are clearly distinguishable, as for the
University of Washington mesoscale ensemble \citep{em05}. 
However, most of the currently used ensemble
prediction systems produce ensembles where some ensemble
members are statistically indistinguishable. Usually, these
exchangeable ensemble members are obtained with the help of 
perturbations of the initial conditions, which is the case for the 51
member European Centre for Medium-Range Weather Forecasts
ensemble \citep{lp} or for the ALADIN-HUNEPS ensemble described in 
Section \ref{sec:sec2}.

Suppose we have \ $M$ \ ensemble members divided into \ $m$ \ exchangeable
groups, where the \ $k$th \ group contains \ $M_k\geq 1$ \ ensemble
members, so \ $\sum_{k=1}^mM_k=M$. \ Further, denote by \ $f_{k,\ell}$
\ the  $\ell$th member of the $k$th group. For this
situation \citet{frg} suggested to use model
\begin{equation}
  \label{eq:eq3.2}
p(x\vert f_{1,1}, \ldots ,f_{1,M_1}, \ldots ,  f_{m,1}, \ldots
,f_{m,M_m} ;\theta_1, \ldots 
,\theta_m):=\sum_{k=1}^m\sum_{\ell=1}^{M_k}\omega _k g_k(x \vert\,
f_{k,\ell}, \theta_k), 
\end{equation}
where ensemble members within a given group have the same weights,
PDFs and parameters.

To simplify notations we give the results and
formulae of this section for model \eqref{eq:eq3.1}, but their
generalization to model 
\eqref{eq:eq3.2} is rather straightforward.

\subsection{Truncated normal model}
  \label{subs:subs3.2}
As it was mentioned in the Introduction, for weather variables such as
temperature, pressure or wind vectors BMA models with normal component
PDFs can be fit reasonably well. However, for modeling wind speed, which can
take only nonnegative values, a skewed distribution is required. A
popular candidate is the Weibull distribution \citep[see
e.g.][]{jhmg}, while \citet{sgr10} considers a BMA model based on
gamma distribution. Here we follow the ideas of \citet{glwga} and
\citet{tg} and for wind speed modeling we employ a truncated normal
distribution with a cut-off 
at zero \ ${\mathcal N}^{\, 0}\big(\mu,\sigma^2\big)$ \ with PDF
\begin{equation}
  \label{eq:eq3.3}
g(x\vert\, \mu,\sigma):=\frac{\frac
  1{\sigma}\varphi\big((x-\mu)/\sigma\big)}{\Phi\big(\mu/\sigma\big)
}, \quad x\geq 0, \qquad \text{and} \qquad g(x\vert\, \mu,\sigma):=0,
\quad \text{otherwise,}
\end{equation}
where \ $\varphi$ \ and \ $\Phi$ \ denote the PDF and the cumulative
distribution function (CDF) of the
standard normal distribution, respectively. The mean \ $\kappa$ \ and
variance \ $\varrho^2$ \ of \
${\mathcal N}^{\, 0}\big(\mu,\sigma^2\big)$ \ are 
\begin{equation}
  \label{eq:eq3.4}
\kappa=\mu
+\frac{\sigma\varphi\big(\mu/\sigma\big)}{\Phi\big(\mu/\sigma\big)}
\qquad \text{and} \qquad \varrho^2=\sigma^2 \left(1-
  \frac{\mu}{\sigma}\frac{\varphi\big(\mu/\sigma\big)}{\Phi
    \big(\mu/\sigma\big)} -\Bigg(\frac{\varphi\big(\mu/\sigma\big)}{\Phi
    \big(\mu/\sigma\big)}\Bigg)^2 
\right),
\end{equation}
respectively.

We assume that location \ $\mu$ \ is a linear function of
the forecasted wind speed that leads to BMA mixture model
\begin{equation}
  \label{eq:eq3.5}
p(x\vert\, f_1, \ldots ,f_M;\alpha_1, \ldots
,\alpha_M; \beta_1, \ldots ,\beta_M; \sigma_1, \ldots
,\sigma_M):=\sum_{k=1}^M\omega _k g(x \vert\, \alpha_k+\beta_k f_k,
\sigma_k),
\end{equation}
where \ $g$ \ is the PDF defined by \eqref{eq:eq3.3}. To simplify the
model, in what follows we also assume \ $\sigma_1=\sigma_2=\ldots
=\sigma_M=:\sigma $. This reduces the number of parameters and makes
computations easier. Furthermore, the BMA gamma model of
\citet{sgr10}, used as a reference, also operates with this restriction.

\subsection{Continuous ranked probability score}
  \label{subs:subs3.3}

Continuous ranked probability score (CRPS) is the most popular scoring
rule for evaluating density forecasts \citep{grjasa,wilks2}. Given a CDF
\ $F(y)$ \ and a real number \ $x$, \ the CRPS is defined as 
\begin{equation}
  \label{eq:eq3.6}
\crps\big(F,x\big):=\int_{-\infty}^{\infty}\big (F(y)-{\mathbbm 
  1}_{\{y \geq x\}}\big )^2{\mathrm d}y={\mathsf E}|X-x|-\frac 12
{\mathsf E}|X-X'|, 
\end{equation}
where \ $X$ \ and \ $X'$ \ are independent random variables with CDF \
$F$ \ and finite first moment. CRPS is a proper scoring rule which is
negatively oriented, i.e. the smaller the better, and it can be
reported in the same unit as the observation. 

Now, if \ $X\sim {\mathcal N}^{\,0}\big(\mu,\sigma^2\big)$ \ then short
calculation shows
\begin{equation*}
S_1(x,\mu,\sigma):={\mathsf E}|X-x|=\Big[\Phi\big(\mu/\sigma\big)\Big]^{-1}
\bigg[A\big(x-\mu,\sigma^2\big)  
 +(x-\mu)\Big(\Phi\big(\mu/\sigma\big)-1\Big)-
\sigma\varphi\big(\mu/\sigma\big)\bigg],   
\end{equation*}
where \ $A\big(\mu,\sigma^2\big):={\mathsf E}|Y|$ \ with \ $Y\sim
{\mathcal N}\big(\mu,\sigma^2\big)$, \ that is 
\begin{equation*}
A\big(\mu,\sigma^2\big)=\mu
\Big(2\Phi\big(\mu/\sigma\big)-1\Big)+2\sigma
\varphi\big(\mu/\sigma\big).
\end{equation*}

Further, if \ $X_1\sim {\mathcal N}^{\, 0}\big(\mu_1,\sigma_1^2\big)$ \ and \
$X_2\sim {\mathcal N}^{\, 0}\big(\mu_2,\sigma_2^2\big)$ \ are independent,
then the PDF of \ $|X_1-X_2|$ \ is
\begin{equation*}
f_{|X_1\!-\!X_2|}(x)\!=\!\Bigg[\sigma_d\Phi\bigg(\frac{\mu_1}{\sigma_1}\bigg)
\Phi\bigg(\frac{\mu_2}{\sigma_2}\bigg)\Bigg]^{-1}\!
\Bigg[\varphi\bigg(\frac{x\!-\!\mu_d}{\sigma_d}\bigg)
\Phi\bigg(\varrho_d\!-\!\frac{\sigma_2 x}{\sigma_1\sigma_d}\bigg) 
\!+\!\varphi\bigg(\frac{x\!+\!\mu_d}{\sigma_d}\bigg)
\Phi\bigg(\varrho_d\!-\!\frac{\sigma_1 x}{\sigma_2\sigma_d}\bigg)\Bigg],
\end{equation*}
where
\begin{equation*}
  \mu_d:=\mu_1-\mu_2, \qquad \sigma_d:=\sqrt{\sigma_1^2+\sigma_2^2}
  \qquad \text{and} \qquad \varrho_d:=\frac {\mu_1\sigma_2^2
    +\mu_2\sigma_1^2}{\sigma_1\sigma_2\sigma_d}. 
\end{equation*}
Hence,
\begin{align*}
S_2(\mu_1,\mu_2,\sigma_1,\sigma_2):={\mathsf
  E}|X_1-X_2|=&\,\Big[\Phi\big(\mu_1/\sigma_1\big) 
\Phi\big(\mu_2/\sigma_2\big)\Big]^{-1} \\ 
&\times \Big[A\big(\mu_1-\mu_2,\sigma_1^2
+\sigma_2^2 \big) -\sqrt{\sigma_1^2+\sigma_2^2}
\,C(\mu_1,\mu_2,\sigma_1,\sigma_2)\Big], \nonumber 
\end{align*}
where the correction term
\begin{equation*}
C(\mu_1,\mu_2,\sigma_1,\sigma_2):=\int_0^{\infty}x
\bigg[\varphi\bigg(x-\frac{\mu_d}{\sigma_d}\bigg) 
\Phi\bigg(\frac{\sigma_2}{\sigma_1}x-\varrho_d\bigg) 
+\varphi\bigg(x+\frac{\mu_d}{\sigma_d}\bigg)
\Phi\bigg(\frac{\sigma_1}{\sigma_2}x-\varrho_d\bigg)\bigg]{\mathrm d}x 
\end{equation*}
can only be evaluated numerically. Now, using \eqref{eq:eq3.6} one can
easily obtain the CRPS corresponding to the CDF \
${\mathcal P}$ \ of mixture model \eqref{eq:eq3.5}, namely
\begin{align*}
\crps\big({\mathcal P},x\big)=\sum _{k=1}^M\omega _k
S_1(x,\alpha_k+\beta_kf_k,\sigma _k)-\frac
12\sum_{k=1}^M\sum_{\ell=1}^M\omega _k\omega _{\ell}
S_2(\alpha_k+\beta_kf_k,\alpha_{\ell}+\beta_{\ell}f_{\ell },\sigma _k,\sigma_{\ell}).
\end{align*}

\subsection{Parameter estimation}
  \label{subs:subs3.4}

Parameters \ $\alpha_k, \ \beta_k, \ \omega_k, \ k=1,2,\ldots M$, \ and
\ $\sigma$ \ are estimated using training data consisting of ensemble
members and verification observations from the preceding \ $n$ \ days
(training period). In
what follows, \ $f_{k,s,t}$ \ denotes the $k$th ensemble member
for location \ $s\in{\mathcal S}$ \ and  time \ $t\in{\mathcal T}$ \
and by \ $x_{s,t}$ \ we denote the corresponding validating
observation.  We consider three approaches which mainly differ in 
estimation of the parameters \ $\alpha_k, \ \beta_k$ \ of location. 

\subsubsection{Naive approach}
  \label{subs:subs3.4.1}
Similarly to the normal BMA model of \citet{rgbp} we 
estimate location parameters  
$\alpha_k$ \ and  \ $\beta_k$ \ with a linear regression of \
$x_{s,t}$ \ on   \ $f_{k,s,t}$ \ over  the time points in the training
period. We call this approach naive since it does not take into
account that for truncated normal distribution  \ ${\mathcal N}^{\,
  0}\big(\mu,\sigma^2\big)$ \  location \ $\mu$ \ does not
equal to the mean \ $\kappa$. \ However, one should remark that with
the increase 
of \ $\mu$ \ the correction term in \eqref{eq:eq3.4} decreases in an
exponential rate. 

To estimate weights \ $\omega_k, \ k=1,2,\ldots ,M$, \ and scale
parameter \ $\sigma$ \  maximum likelihood method is applied using
again training data. Under the assumption of independence of forecast
errors in space and time the log-likelihood function corresponding to
model \eqref{eq:eq3.5} equals
\begin{equation}
  \label{eq:eq3.7}
\ell (\omega_1,\ldots ,\omega_M;\sigma)=\sum_{s,t}\log \left[ 
  \sum_{k=1}^M \omega_k g\big(x_{s,t}\vert \, \alpha_k+\beta_k f_{k,s,t},
    \sigma \big) \right],
\end{equation}
where the first summation is over all locations \ $s\in{\mathcal S}$ \
and time points \ $t$ \ from the training period containing \ $N$ \
terms \ ($N$ \ distinct values of \ $(s,t)$). 

The log-likelihood function \eqref{eq:eq3.7} is too complicated to be
maximized analytically, so we find its maximum using the truncated data EM
algorithm of \citet{ls12}. Similarly to the traditional EM algorithm
for mixtures \citep{dlr,mclk} we introduce latent indicator variables
\ $z_{k,s,t}$ \ taking values one or zero according as whether \
$x_{s,t}$ \ comes from the $k$th component PDF or not. The  complete
data log-likelihood corresponding to observations and indicator
variables equals
\begin{equation}
  \label{eq:eq3.8}
\ell_C (\omega_1,\ldots ,\omega_M; \sigma)
=\sum_{s,t} \sum_{k=1}^M
z_{k,s,t} \bigg[\log(\omega_k) + \log \Big(
  g\big(x_{s,t}\vert \, \alpha_k+\beta_k f_{k,s,t},
    \sigma \big)\Big)\bigg]. 
\end{equation}

The EM algorithm alternates between an
expectation (E) step and a maximization (M) step until convergence.
It starts with initial values \ $\omega_k^{(0)}, \ k=1,2,\ldots ,M,$ \
and  \ $\sigma^{(0)}$ \ of the parameters. In the E step the latent
variables are estimated using the conditional expectation of the
complete log-likelihood on the observed data, while in the M step
the parameter estimates are updated by maximizing \ $\ell_C$ \ with
the current values of the latent variables plugged in. 

For the truncated normal mixture model given by \eqref{eq:eq3.3} and
\eqref{eq:eq3.5} the E step is,
\begin{equation}
  \label{eq:eq3.9}
z_{k,s,t}^{(j+1)}:=\frac {\omega_k^{(j)}g\big(x_{s,t}\vert \,
  \alpha_k+\beta_k f_{k,s,t}, \sigma^{(j)} \big)}{\sum
  _{i=1}^M\omega_i^{(j)}g\big(x_{s,t}\vert \, 
  \alpha_i+\beta_i f_{i,s,t}, \sigma^{(j)} \big)},
\end{equation}
where the superscript refers to the actual iteration. The first part
of the M step is obviously
\begin{equation}
  \label{eq:eq3.10}
\omega_k^{(j+1)}:=\frac 1N\sum_{s,t}z_{k,s,t}^{(j+1)}, 
\end{equation}
while the second part can be derived from equation \ $\frac{\partial
  \ell_C}{\partial \sigma} =0$. \ However, in our case this equation
is nonlinear and since it cannot be solved for \ $\sigma$, \ we
suggest iteration step
\begin{align}
  \label{eq:eq3.11}
\sigma^{2(j+1)}=\frac 1N &\sum_{s,t}\sum_{k=1}^M z_{k,s,t}^{(j+1)}
\big (x_{s,t}-\alpha_k-\beta_kf_{k,s,t}\big)^2 \\
&+\frac {\sigma^{(j)}}N
\sum_{s,t}\sum_{k=1}^M z_{k,s,t}^{(j+1)}
\big(\alpha_k+\beta_kf_{k,s,t}\big) \frac {\varphi
  \big((\alpha_k+\beta_kf_{k,s,t})/\sigma^{(j)} \big)}{\Phi
  \big((\alpha_k+\beta_kf_{k,s,t})/\sigma^{(j)}\big)}. \nonumber
\end{align}
Observe that the EM algorithm presented here differs from the corresponding
algorithm of \citet{rgbp} only in the  second term of
\eqref{eq:eq3.11}. 

\subsubsection{Mean corrected approach}
  \label{subs:subs3.4.2}

In this approach we assume that the  means of the component PDFs are also 
linear functions  of form \ $a_k+b_kf_{k,s,t}$ \  of the forecasted
wind speed and the linear regression of \ $x_{s,t}$ \ on   \
$f_{k,s,t}$ \ is now used to estimate \ $a_k$ \ and \ $b_k, \
k=1,2,\ldots ,M$. \ Instead of \eqref{eq:eq3.8} we consider the complete
data log-likelihood
\begin{equation*}
\ell_C (\omega_1,\ldots ,\omega_M, \sigma)=\sum_{s,t} \sum_{k=1}^M
z_{k,s,t} \bigg[\log(\omega_k) + \log \Big(
  g\big(x_{s,t}\vert \, \mu_{k,s,t},
    \sigma \big)\Big)\bigg],
\end{equation*}
where the initial guess for the location parameter \ $\mu_{k,s,t}$ \
of the $k$th component PDF at \ $(s,t)$ \ is \ $\mu_{k,s,t}^{(0)}:=a_k+b_kf_{k,s,t}$.

Now, E step \eqref{eq:eq3.9} is replaced by
\begin{equation}
  \label{eq:eq3.12}
z_{k,s,t}^{(j+1)}:=\frac {\omega_k^{(j)}g\big(x_{s,t}\vert \,
  \mu_{k,s,t}^{(j)}, \sigma^{(j)} \big)}{\sum
  _{i=1}^M\omega_i^{(j)}g\big(x_{s,t}\vert \, 
  \mu_{i,s,t}^{(j)}, \sigma^{(j)} \big)},
\end{equation}
the first part \eqref{eq:eq3.10} of the M step remains valid, while
\begin{equation}
   \label{eq:eq3.13}
  \mu_{k,s,t}^{(j+1)}:=\mu_{k,s,t}^{(0)}-\sigma^{(j)}\frac {\varphi
  \Big(\mu_{k,s,t}^{(j)}/\sigma^{(j)} \Big)}{\Phi
  \Big(\mu_{k,s,t}^{(j)}/\sigma^{(j)}\Big)}
\end{equation}
and
\begin{equation}
  \label{eq:eq3.14}
\sigma^{2(j+1)}=\frac 1N \sum_{s,t}\sum_{k=1}^M z_{k,s,t}^{(j+1)}
\Big (x_{s,t}- \mu_{k,s,t}^{(j+1)}\Big)^2 
+\frac {\sigma^{(j)}}N
\sum_{s,t}\sum_{k=1}^M z_{k,s,t}^{(j+1)}  \mu_{k,s,t}^{(j+1)} 
\frac {\varphi
  \Big(\mu_{k,s,t}^{(j+1)}/\sigma^{(j)} \Big)}{\Phi
  \Big(\mu_{k,s,t}^{(j+1)}/\sigma^{(j)}\Big)}
\end{equation}
substitute iteration step \eqref{eq:eq3.11}. Observe that
\eqref{eq:eq3.13} is responsible for the mean correction in the first
term of \eqref{eq:eq3.4}. Finally, after the EM algorithm stops,
parameters \ $\alpha_k$ \ and \ $\beta_k$ \ are estimated with a
linear regression of the final value of \ $\mu_{k,s,t}$ \ on \ $f_{k,s,t}$.

\subsubsection{Full maximum likelihood estimation}
  \label{subs:subs3.4.3}

In the previous two cases the estimates of location parameters \
$\alpha_k$ \ and \ $\beta_k, \ 
k=1,2,\ldots ,M$, \  are obtained separately from the weights and
the scale parameter. Here we present a method where all parameters are
estimated with ML method using the
complete data log-likelihood defined by \eqref{eq:eq3.8}. Obviously,
in this case \ $\ell_C$ \ is considered as a function of \ $\alpha_k, \
\beta_k, \ \omega_k, \ k=1,2,\ldots ,M$, and \ $\sigma$. \ The initial
guesses  \ $\alpha_k^{(0)}$ \ and \
$\beta_k^{(0)}$ \ for these parameters are the corresponding mean
coefficients 
obtained from regressing of \ $x_{s,t}$ \ on \
$f_{k,s,t}$ \ over the time points in the training period. In this way
the initial value \
$\mu_{k,s,t}^{(0)}:=\alpha_k^{(0)}+\beta_k^{(0)}f_{k,s,t}$ \ of the
location parameter is exactly the estimated mean of the $k$th
component PDF.

The E step remains the same as in the mean corrected approach, that is
estimate \ $z_{k,s,t}^{(j+1)}$ \ of the latent
variable \ $ z_{k,s,t}$ \ is given by \eqref{eq:eq3.12}. The first
part \eqref{eq:eq3.10} of the M step remains valid again, while
equations \ $\frac {\partial \ell_C}{\partial \alpha_k}$ \ and  \
$\frac {\partial \ell_C}{\partial \beta_k}$ \ result in iterations
\begin{align*}
\alpha_k^{(j+1)}:=&\,\bigg[\sum_{s,t}z_{k,s,t}^{(j+1)}\bigg]^{-1}\left[\sum_{s,t}
z_{k,s,t}^{(j+1)} \left(
\Big (x_{s,t}- \beta_k^{(j)}f_{k,s,t}\Big)
-\sigma^{(j)} \frac{\varphi
  \Big(\mu_{k,s,t}^{(j)}/\sigma^{(j)} \Big)}{ \Phi
  \Big(\mu_{k,s,t}^{(j)}/\sigma^{(j)}\Big)}\right)\right], \\
\beta_k^{(j+1)}:=&\,\bigg[\sum_{s,t}z_{k,s,t}^{(j+1)}f_{k,s,t}^2\bigg]^{-1}
\left[\sum_{s,t} z_{k,s,t}^{(j+1)} f_{k,s,t} \left(
\Big (x_{s,t}- \alpha_k^{(j+1)}\Big)
-\sigma^{(j)} \frac{\varphi
  \Big(\mu_{k,s,t}^{(j)}/\sigma^{(j)} \Big)}{ \Phi
  \Big(\mu_{k,s,t}^{(j)}/\sigma^{(j)}\Big)}\right)\right].
\end{align*}
Mean correction \eqref{eq:eq3.13} takes form
\begin{equation}
   \label{eq:eq3.15}
  \mu_{k,s,t}^{(j+1)}:=\mu_{k,s,t}^{(0)}-\sigma^{(j)}\frac {\varphi
  \Big(\big(\alpha_k^{(j+1)}+\beta_k^{(j+1)}f_{k,s,t}\big)/\sigma^{(j)} \Big)}{\Phi
  \Big(\big(\alpha_k^{(j+1)}+\beta_k^{(j+1)}f_{k,s,t}\big)/\sigma^{(j)}\Big)},
\end{equation}
while the estimate of variance is updated using \eqref{eq:eq3.14}.

\section{Results}
  \label{sec:sec4}

As was mentioned in the Introduction, the performance of BMA
model \eqref{eq:eq3.5} is tested on the ALADIN-HUNEPS ensemble of
HMS.  We consider all three parameter estimating methods of Subsection 
\ref{subs:subs3.4} and use the same data base as in \citet{bhn1}, 
where the authors calibrated the raw ensemble with the help of the BMA
gamma model of \citet{sgr10} considering a training period of 28 calendar
days. Here we apply the same training
period length which allows direct comparison of the two BMA methods.  
In this way ensemble members, validating observations and BMA models
are available for 146 calendar days (on 20.11.2010 all ensemble
members are missing).  

\subsection{Models and diagnostics}
  \label{subs:subs4.1}
Using the ideas of \citet{bhn1} we consider two different 
groupings of ensemble members. In the first case we have two
exchangeable groups. One contains the control denoted by \ $f_c$ \
while in the other are 10 ensemble members
corresponding to the different perturbed initial conditions denoted by
\ $f_{p,1},\ldots ,f_{p,10}$. \  This leads us to model
\begin{align} 
  \label{eq:eq4.1}
p\big(x |\, f_c\,f_{p,1},\ldots ,
f_{p,10};\alpha_c,\alpha_p;\beta_c,\beta_p;\sigma\big) = &\,\omega
g\big(x|\,\alpha_c+\beta_c f_c, \sigma\big) \\ 
&+\frac {1-\omega}{10}
\sum_{\ell=1}^{10} 
g\big(x|\,\alpha_p+\beta_pf_{p,\ell},\sigma\big),  \nonumber
\end{align} 
where \ $\omega\in [0,1]$, \ and \  $g$ \ is defined by
\eqref{eq:eq3.3}.

In the second case the odd and even numbered exchangeable ensemble
members form two separate groups
$\{f_{p,1}, \ 
f_{p,3}, \ f_{p,5}, \ f_{p,7}, \ f_{p,9}\}$ \ and \ $\{f_{p,2}, \
f_{p,4}, \ f_{p,6}, \ f_{p,8}, \ f_{p,10}\}$, \ respectively,  which idea is
justified by the method their initial conditions
are generated. To get them only five perturbations are calculated and
then they are added to (odd numbered members) and
subtracted from (even numbered members) the unperturbed
initial conditions \citep{horanyi,bhn1,bhn2}. In
this way we obtain the following PDF for the forecasted wind speed:
\begin{align}
  \label{eq:eq4.2}
q\big(x |\, f_c,f_{p,1},\ldots ,
f_{p,10};&\,\alpha_c, \alpha_o, \alpha_e; \beta_c,\beta_o,\beta_e;\sigma
\big)= \omega_c 
g\big(x|\,\alpha_c+f_c\beta_c,\sigma\big) \\ &+ 
\sum_{\ell=1}^{5} \Big(\omega_o g\big(x|\,\alpha_o+\beta_of_{p,2\ell
  -1},\sigma\big)+  
\omega_e g\big(x|\,\alpha_e+\beta_ef_{p,2\ell},\sigma\big)\Big), \nonumber 
\end{align} 
where for weights \ $\omega_c,\omega_o,\omega_e\in[0,1]$ \ we have  \
$\omega_c+5\omega_o+5\omega_e=1$.

\begin{figure}[t]
\begin{center}
\leavevmode
\epsfig{file=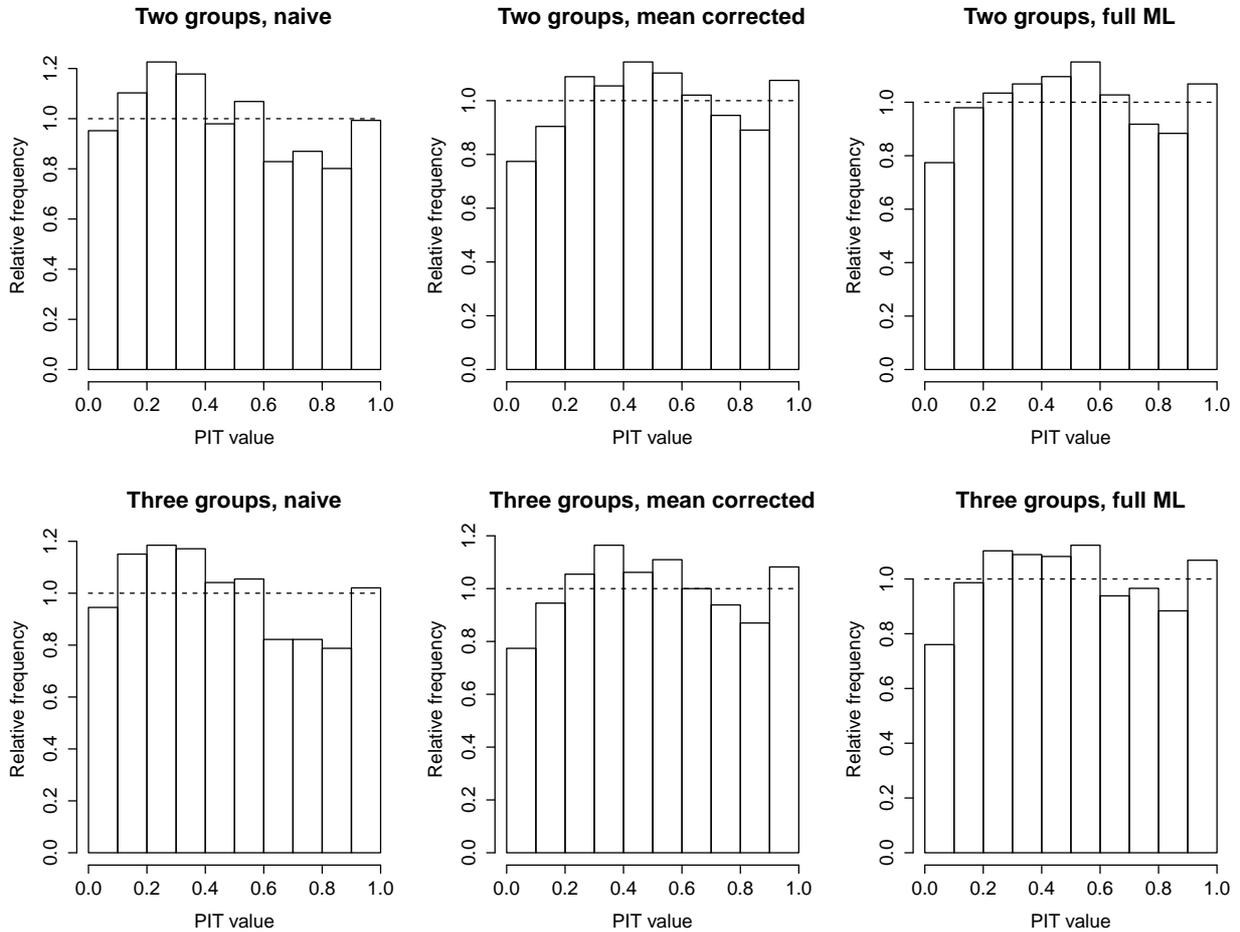,height=16.5cm, angle=-90}
\caption{PIT histograms for BMA post-processed forecasts using two-group
  \eqref{eq:eq4.1} and three-group \eqref{eq:eq4.2} models.} 
\label{fig:fig2}
\end{center}
\end{figure} 
In order to check the overall performance of the probabilistic forecasts
(based on models \eqref{eq:eq4.1} and \eqref{eq:eq4.2}) in terms of a
probability distribution function, the mean CRPS (the 
average of the CRPS values 
of the predictive CDFs and corresponding validating observations taken
over all locations and time points considered) 
and the coverage and average widths of 66.7\,\% and 90\,\%
central prediction intervals are computed and compared for both BMA
methods (truncated normal and gamma) and raw ensemble. In the latter
case, the ensemble of 
forecasts corresponding to a given location and time is considered as a
statistical sample, the empirical CDF of the ensemble replaces the
predictive CDF and the sample quantiles are calculated according 
to \citet[Definition 7]{hf}.  We remark that the
coverage of a \ $(1-\alpha)100 \,\%, \ \alpha \in (0,1)$ \ central prediction
interval is the proportion 
of validating observations located between the lower and upper \
$\alpha/2$ \ quantiles of the predictive distribution. For a
calibrated predictive PDF this value should be around \ $(1-\alpha)100
\,\%$.   
Additionally, the BMA and ensemble medians are  considered as point
forecasts, which are evaluated with the use 
of mean absolute errors (MAE) and root mean square errors (RMSE). 

\subsection{Verification results of BMA post-processing}
 \label{subs:subs4.2}

To get a first insight about the calibration of BMA post-processed
forecasts we consider probability integral transform (PIT)
histograms. The PIT is 
the value of the predictive cumulative distribution evaluated at
the verifying observations \citep{rgbp}, which is providing a good
measure about the possible improvements of the under-dispersive
character of the raw ensemble.  The closer the histogram is to the
uniform distribution, the better the calibration is. In Figure
\ref{fig:fig2} the PIT histograms corresponding to all three parameter
estimating methods and to both BMA models
\eqref{eq:eq4.1} and \eqref{eq:eq4.2} are displayed. A comparison to
the verification rank histogram of the raw ensemble (see  Figure
\ref{fig:fig1}) shows that post-processing significantly improves the
statistical calibration of the forecasts. Further, on the basis of 
significance levels of Kolmogorov-Smirnov tests  given in
Table \ref{tab:tab1}, for both models \eqref{eq:eq4.1} and
\eqref{eq:eq4.2} one can accept the uniformity of PIT values
corresponding to truncated normal BMA model with full maximum likelihood
parameter estimation method. However, one should remark that for the
three group model mean corrected parameter estimation also yields
acceptable PIT values.

\begin{table}[t!]
\begin{center}
\begin{tabular}{|l|c|c|c|c|} \hline
\multicolumn{1}{|l|}{}&\multicolumn{3}{|c|}{Truncated normal
  BMA}&Gamma BMA \\ \cline{2-4}
&naive&mean corr.&full ML& \\ \hline
Two groups&$2.42\times 10^{-4}$&$1.79\times 10^{-2}$&$0.13$&$2.22\times 10^{-2}$\\
Three groups&$1.37\times 10^{-4}$&$5.56\times
10^{-2}$&$0.18$&$1.87\times 10^{-2}$ \\ \hline  
\end{tabular} 
\caption{Significance levels of Kolmogorov-Smirnov tests for
  uniformity of PIT 
  values corresponding to two- and three-group models.} \label{tab:tab1}      
\end{center}
\end{table} 
 
In Table \ref{tab:tab2} scores for different
probabilistic forecasts are given together with the average width and
coverage of $66.7\,\%$ and 
$90.0\,\%$ central prediction intervals. Verification measures of
probabilistic forecasts and point forecasts calculated using truncated
normal BMA models \eqref{eq:eq4.1} and \eqref{eq:eq4.2} are compared
to the corresponding measures calculated for the raw ensemble and
applying gamma BMA post-processing \citep{bhn1}. Compared to the raw
ensemble all 
BMA post-processed forecasts show a significant decrease in all there
verification scores considered. Further, as the listed CRPS, MAE and
RMSE values show, the accuracy of the truncated
normal BMA probabilistic and point forecasts is better than the
accuracy of the gamma BMA 
ones. 

Concerning calibration, one can observe
that the coverage of both BMA central prediction intervals
are rather close to the correct coverage for all models considered, while
the coverage of the central prediction intervals calculated
from the raw ensemble are quite poor. This shows that BMA
post-processing greatly improves calibration.
Further, the truncated normal BMA models yields slightly sharper
predictions than the gamma BMA forecasts and  one can also observe
that the three-group model slightly outperforms the two-group one.

\begin{table}[t!]
\begin{center}
{
\begin{tabular}{|l|l|l|c|c|c|c|c|c|c|} \hline
\multicolumn{3}{|l|}{}&Mean&MAE&RMSE&\multicolumn{2}{|c|}{Average
  width}&\multicolumn{2}{|c|}{Coverage (\%)} \\ \cline{7-10}
\multicolumn{3}{|c|}{Forecast}&CRPS&&&66.7\%
&90.0\% &66.7\% &90.0\% \\ \hline
&Trunc.&naive&$0.7225$&$1.0631$&$1.3800$&$2.5738$&$4.2850$&$67.81$&$89.79$ \\
\cline{3-10} 
Two&normal&mean
c.&$0.7062$&$1.0520$&$1.3784$&$2.6175$&$4.3420$&$69.86$&$90.55$ \\
\cline{3-10}
groups&BMA&full
ML&$0.7071$&$1.0518$&$1.3786$&$2.6029$&$4.3222$&$69.38$&$90.14$ \\
\cline{2-10} 
&\multicolumn{2}{|c|}{Gamma
  BMA}&$0.7577$&$1.0678$&$1.4213$&$2.6359$&$4.5297$&$68.08$&$90.34$ \\
\hline 
&Trunc.&naive&$0.7213$&$1.0612$&$1.3771$&$2.5645$&$4.2709$&$67.26$&$89.86$ \\
\cline{3-10} 
Three&normal&mean
c.&$0.7042$&$1.0480$&$1.3737$&$2.6043$&$4.3195$&$69.38$&$90.34$ \\
\cline{3-10}
groups&BMA&full
ML&$0.7044$&$1.0485$&$1.3739$&$2.5948$&$4.3073$&$68.84$&$90.14$ \\
\cline{2-10} 
&\multicolumn{2}{|c|}{Gamma
  BMA}&$0.7556$&$1.0643$&$1.4153$&$2.6153$&$4.4931$&$68.36$&$90.21$ \\
\hline 
\multicolumn{3}{|c|}{Raw
  ensemble}&$0.8599$&$1.1215$&$1.4634$&$1.4388$&$2.2001$&$38.70$&$55.14$ \\
\hline 
\end{tabular} 
}
\caption{Mean CRPS of probabilistic, MAE and RMSE of
median forecasts, average width and coverage of $66.7\,\%$ and
$90.0\,\%$ central prediction intervals.} \label{tab:tab2}
\end{center}
\end{table}

Finally, we remark that mean correction step \eqref{eq:eq3.15}
in full ML parameter estimation method seems essential. Running the algorithm
without it (that is $ \mu_{k,s,t}^{(j+1)}:=\alpha_k^{(j+1)} 
+\beta_k^{(j+1)}f_{k,s,t}$) e.g. for the three group model yields 
smaller CRPS ($0.7024$) but larger MAE and RMSE values ($1.0499$ 
and $1.3900$) and wider central prediction intervals. Moreover,
in this case the PIT values do not fit the uniform distribution.

\section{Discussion}
  \label{sec:sec5}

We introduced a new BMA model for post-processing ensemble forecasts of
wind speed providing a predictive PDF which is a mixture of normal
distributions truncated from below at zero. The model was tested on
the 11 member ALADIN-HUNEPS ensemble of the HMS using two different
BMA models. One assumes two groups of exchangeable members (control
and forecasts from perturbed initial conditions), while the other considers
three (control and forecasts from perturbed initial conditions
with positive and negative perturbations). For
both models a 28 day training period was used and three types of
parameter estimation: a naive and two more sophisticated one with mean
correction and full maximum likelihood estimation. The latter resulted
PIT values which perfectly fit the 
uniform distribution both for the two- and for the three-group
model. The comparison of the raw ensemble and of the truncated 
normal BMA forecasts shows that the mean CRPS values of BMA
post-processed forecasts are considerably lower than the mean CRPS of
the raw ensemble. Furthermore, the MAE and RMSE values of BMA median forecasts
are also lower than the MAEs and
RMSEs of the ensemble median. The calibration of BMA forecasts is
nearly perfect as the coverage of the $66.7\,\%$ and $90.0\,\%$
prediction intervals are very close to the nominal levels.  From the
three competing parameter estimation methods the overall performance of the
full ML estimation seems to be the best.

Compared to the performance of gamma BMA model of \citet{sgr10} for
wind speed data investigated in \citet{bhn1} one can observe that truncated
normal BMA post-processing yields slightly lower CRPS, MAE and RMSE values
and narrower central prediction intervals. However, the great
advantage of the model presented here 
appears in the speed of parameter estimation. The EM
algorithm for estimating the 
weights and distribution parameters of the truncated normal BMA model uses
closed formulae while in case of the gamma BMA model a numerical
optimization is required. Running the {\tt ensembleBMA} package of R
\citep{frgs, frgsb} on a PC under a 64bit Window 7 operating system
(Intel Core i5-3470 CPU, 3.40 GHz, 4 cores, 8 Gb RAM) the estimation
of parameters of the three-group model
e.g. for 25.03.2011 using training period of 28 calendar days, for the
gamma BMA model took $10.23$ seconds, while for the truncated normal with
naive, mean corrected and full ML parameter estimation methods, $6.86$
s, $6.83$ s and
$10.41$ s, respectively. The difference is more convincing when we perform
modeling for all possible 148 calendar days. The corresponding running times
were $2716.98$ s for the gamma BMA and $797.89$ s, $918.16$ s and
$1245.5$ s for the truncated normal BMA models. In this way the truncated
normal BMA model outperforms the traditional gamma BMA model both in
accuracy and calibration of forecasts and in computation time.

\bigskip
\noindent
{\bf Acknowledgments.} \  \ Research was supported by 
the Hungarian  Scientific Research Fund under Grant No. OTKA NK101680
and by the T\'AMOP-4.2.2.C-11/1/KONV-2012-0001 
project. The project has been supported by the European Union,
co-financed by the European Social Fund. 
The author is indebted to Tilmann
Gneiting and Andr\'as Hor\'anyi for their useful suggestions and
remarks and to M\'at\'e Mile and  
Mih\'aly Sz\H ucs from the HMS for providing the data.

\end{document}